\documentclass[10pt]{article}
\PassOptionsToPackage{final}{graphicx}
\usepackage{graphicx}
\usepackage[T1]{fontenc}
\usepackage{lmodern}
\usepackage{microtype}
\usepackage{amsmath, amssymb}
\usepackage{hyperref}
\hypersetup{
	colorlinks=true,
	linkcolor=blue,
	filecolor=magenta,      
	urlcolor=cyan}

\begin{document}
\title{Exploring the energy landscape of the logarithmic potential: local minima and stationary states}
\author{Paolo Amore \\
\small Facultad de Ciencias, CUICBAS, Universidad de Colima,\\
\small Bernal D\'{i}az del Castillo 340, Colima, Colima, Mexico \\
\small paolo@ucol.mx  \\
\and
Victor Figueroa \\
\small Facultad de Ciencias, Universidad de Colima,\\
\small Bernal D\'{i}az del Castillo 340, Colima, Colima, Mexico \\
\small vfigueroa6@ucol.mx  \\
\and
Raymundo Ramos \\
\small  Quantum Universe Center, Korea Institute for Advanced Study, Seoul 02455, Korea  \\
\small raramos@kias.re.kr  \\
}

\maketitle
\begin{abstract}
We have performed a detailed exploration of the energy landscape for configurations of points on the sphere, interacting via the logarithmic potential,  and corresponding to 
 local minima of the total energy, up to $N = 160$.  The growth of  $N_{\rm conf}$  (number of distinct configurations) is exponential, as for the Thomson problem,
although  weaker. Using the techniques described in our previous paper~\cite{Amore25} we have also explored the solution landscape of this problem for $N \leq 24$, and found that the number of stationary states is growing exponentially. 
\end{abstract}

\maketitle

\section{Introduction}
\label{intro}

In this paper we extend the previous analysis by two of us, ref.~\cite{Amore25}, for the Thomson problem to the logarithmic potential, which can be regarded as a special case of the Riesz potential $V = 1/r^s$, in the limit $s \rightarrow 0$. Given $N$ points on a sphere, interacting via a repulsive potential $V$, the basic questions one wants to answer are: how do the points distribute on the surface of the sphere, when reaching a position of equilibrium? how many local minima does the system have? what is the global minimum of the problem at fixed $N$? 

These questions, however, cannot be answered in general, unless $N$ is sufficiently small, and empirical (partial) answers can be found only by performing numerical experiments. For instance, a useful way to appreciate the distribution of points on the sphere is to visualize the corresponding Voronoi diagram, which, for the sphere, must correspond to a total topological charge $Q=12$, because of Euler's theorem: at given $N$, the energy of the configuration depends on the way the topological defects distribute on the sphere, while maintaining $Q$ fixed, and typically one observes more "exotic" defects for configurations of higher energy; for larger $N$, it is also seen that even the low energy configurations acquire a more complex defect structure. 

The second and third questions, on the other hand, are clearly related as finding {\sl all} possible local minima for a given $N$ implies the possibility of identifying the global minimum: however, these problems have no simple solution, both because the number of local minima grows exponentially with $N$, 
and because there may be configurations that are very difficult to reach.

For the case of the Thomson problem, the effort of determining the growth of the number of configurations with $N$ has been done in a series of paper, starting with the pioneering work of Erber and Hockney~\cite{Erber91,Erber95,Erber97}, and  more recently by Calef and collaborators~\cite{Calef15}, by Mehta and collaborators~\cite{Mehta16} and by Amore and collaborators~\cite{Amore25}. While all these papers agree in stating that  the number of configurations, $N_{\rm conf}$, grows exponentially with $N$, their estimates differ because of the different number of actual configurations found. For completeness, it is worth mentioning that the main goal of Ref.~\cite{Mehta16} was studying the energy landscape for the Thomson problem at selected values of $N$, and not systematically finding all configurations up to a maximum value of $N$: however, the results obtained in that paper at selected values of $N$ drastically improved previous 
calculations for those value. Our analysis for the Thomson problem lead us to discover a much larger number of configurations than previously documented, while at the same time quantifying some aspects of the problem that had not been studied before: in fact we found that the average and minimal energy gaps for configurations with the same number of particles, for instance,  decrease exponentially with $N$, whereas the energy interval over which the configurations 
of same $N$ are distributed grows almost linearly with $N$. Such behavior poses a tremendous technical challenge to this study for values of $N$ large enough so that the average energy separation will be below round--off error.  

The purpose of the present paper is to extend the techniques described in ref.~\cite{Amore25} to systematically explore the energy landscape for the logarithmic potential, for $N \leq 160$ and, additionally, to study the solution landscape (i.e. the landscape for stationary states) for this problem for $N \leq 24$.
The only analysis we are aware of on the growth of $N_{\rm conf}$ with $N$ is contained in ref.~\cite{Calef15} (which also considered the Riesz potentials for $s=2$ and $s=3$). 

Candidates to be the global minima for this potential have been obtained  in refs.~\cite{Bergersen94} and ~\cite{Saff94}: in particular Bergersen et al.~\cite{Bergersen94} pointed out a special property of the equilibrium configurations for the log potential, that all  have vanishing dipole moment, 
regardless of $N$ (this property has also later been proved for larger dimensions ~\cite{Dragnev02}).

This potential is also at the center of the seventh problem of Smale's list~\cite{Smale98}, which asks whether it is possible to algorithmically find configurations of points on the unit sphere that are sufficiently close to the global minimum of the total energy, for particles interacting via a log potential. The problem is still open, see  \cite{Beltran11}. 
  
In this paper we want to address two different aspects in connection to the logarithmic potential: first, to identify the majority of local minima for this problem, for $N \leq 160$, and use these results to obtain an estimate of the exponential growth of $N_{\rm conf}$ with $N$;
second, we wish to carry out an exploration of the solution landscape, for $N \leq 24$, in analogy with what we have done in ref.~\cite{Amore25} for the Thomson problem, with the goal of identifying the majority of stationary states for these values of $N$. The approach of ref.~\cite{Amore25} indeed allows us to transform this problem into an alternative minimization problem, in terms of a suitably defined function.  It is worth saying that, as for the case of the Thomson problem, no previous calculation of this kind has been done before.

Reaching these two goals has implied a  massive amount of numerical experiments, with  stringent requirements for precision.

The paper is organized as follows: in section \ref{sec:num} we describe the different computational strategies that we have used to attack these problems; in section \ref{sec:results} we report the numerical results (the subsections \ref{sec:loc_min} and \ref{sec:stat_states}  contain the cases of local minima and stationary states, respectively). Finally in section \ref{sec:concl} we draw our conclusions.

\section{Computational strategies}
\label{sec:num}

Most of the computational strategies used in the present paper have been described in  ref.~\cite{Amore25}, so that we refer the 
reader to our previous paper. 

The only improvement we have implemented here concerns the {\sl upgrade} and {\sl downgrade} methods: we have not modified the methods themselves, which work the same, but we have devised an approach that uses both on a given range of  values of $N$. We have named this method as "bouncing method" and it works
in the following way:

\begin{itemize}
\item select a interval $N_{\rm min} \leq N \leq N_{\rm max}$ and pick an initial value $N_i$ (typically either $N_{\rm min}$ or $N_{\rm max}$)
\item  at $N_i$ start applying either the upgrade or downgrade method, thus forming configurations for $N_i +1$ or $N_i-1$ (if $N_i = N_{\rm min}$ the upgrade method has to be applied and the downgrade method for $N_i = N_{\rm max}$ );
\item  once the upgrade or downgrade method has finished, the configurations for $N_i+1$ (or $N_i-1$) that had not been upgraded (downgraded) before can be upgraded  (downgraded) as in the previous step, forming configurations for $N_i+2$ ($N_i-2$);
\item iterate these steps until reaching one extreme of the interval, reversing the direction at this value (for instance, reaching $N_{\rm max}$ requires switching from the upgrade to the downgrade method);
\item stop when no new configurations are found;
\end{itemize}

This approach takes advantage of the {\sl deterministic} nature of the {\sl upgrade} and {\sl downgrade} methods and works only on the subset of configurations 
that have not yet been upgraded or downgraded, thus limiting the numerical workload. This method proved particularly useful for the larger values of $N$, for which typically there are several thousands of independent solutions.

\section{Numerical results}
\label{sec:results}

In this section we present the numerical results obtained using the computational approaches described in the previous section. 
These results are available for download at Zenodo.

\subsection{Local minima}
\label{sec:loc_min}

The only other paper where an exploration of the energy landscape for the logarithmic potential has been performed is ref.~\cite{Calef15},  for $100 \leq  N \leq 180$.  It is unfortunate that the number of observed configurations is not reported in that paper, so that 
a comparison is only possible using the exponential fit obtained by the authors or by estimating the approximate values of $N_{\rm conf}$ directly from their plot~\footnote{Notice however that the formula for the estimated number of configurations, $M(N,s)$,  at pag. 250 of \cite{Calef15}
contains a typo and it should probably read $M(N,s) = C_s \ e^{e_s N}$.}. 

If we focus on $N=160$,  the largest $N$ explored in our analysis, we see that Calef et al. found less than $1000$ configurations, while estimating 
less than $3000$ configurations. Our exploration, on the other hand, has lead us to find $14142$ independent configurations, which is a lower bound to the true number of independent configurations.

These configurations emerge from a search that produces tens or even hundreds of thousand of ansatzes, that are minimized with high precision 
and then compared to eliminate redundant configurations.

As we have described in ref.~\cite{Amore25} the identification of different configurations relies on a series of steps:
\begin{itemize}
\item a comparison of the energies: one has to make sure that the configurations have converged to local minima of the problem within a distance small enough so that any difference in energy is not due to lack of convergence;  if $N$ is large enough, however, this criteria is inconclusive, as two configurations that are truly inequivalent may be separated by a gap which is smaller that the round--off error in double precision;
\item a comparison of two configurations relying on a direct comparison of the individual energies of each point;
\item a comparison based on the standard test of ref.~\cite{Wales04} (see also algorithm 1 of ref.~\cite{Calef15});
\end{itemize}

It is also essential that each of the configurations obtained in the numerical calculation be sufficiently precise:  taking as an example the case $N=160$, in particular, we find that the largest norm of the gradient in the population of $14142$ configurations to be $|\nabla V|_{\rm max} = 1.05 \times 10^{-13}$, whereas the largest norm of the center of mass to be even smaller $|{\bf R}_{\rm cm} |= 1.26 \times 10^{-14}$.
As an additional precaution,  for each single configuration that "survived" the filtering  we have calculated the hessian matrix and explicitly verified that its eigenvalues are non--negative. As the reader can guess, the exploration and filtering of local minima for $N \leq 160$ that we have performed is a  demanding numerical task!

\begin{figure}
\begin{center}
\includegraphics[width=10cm]{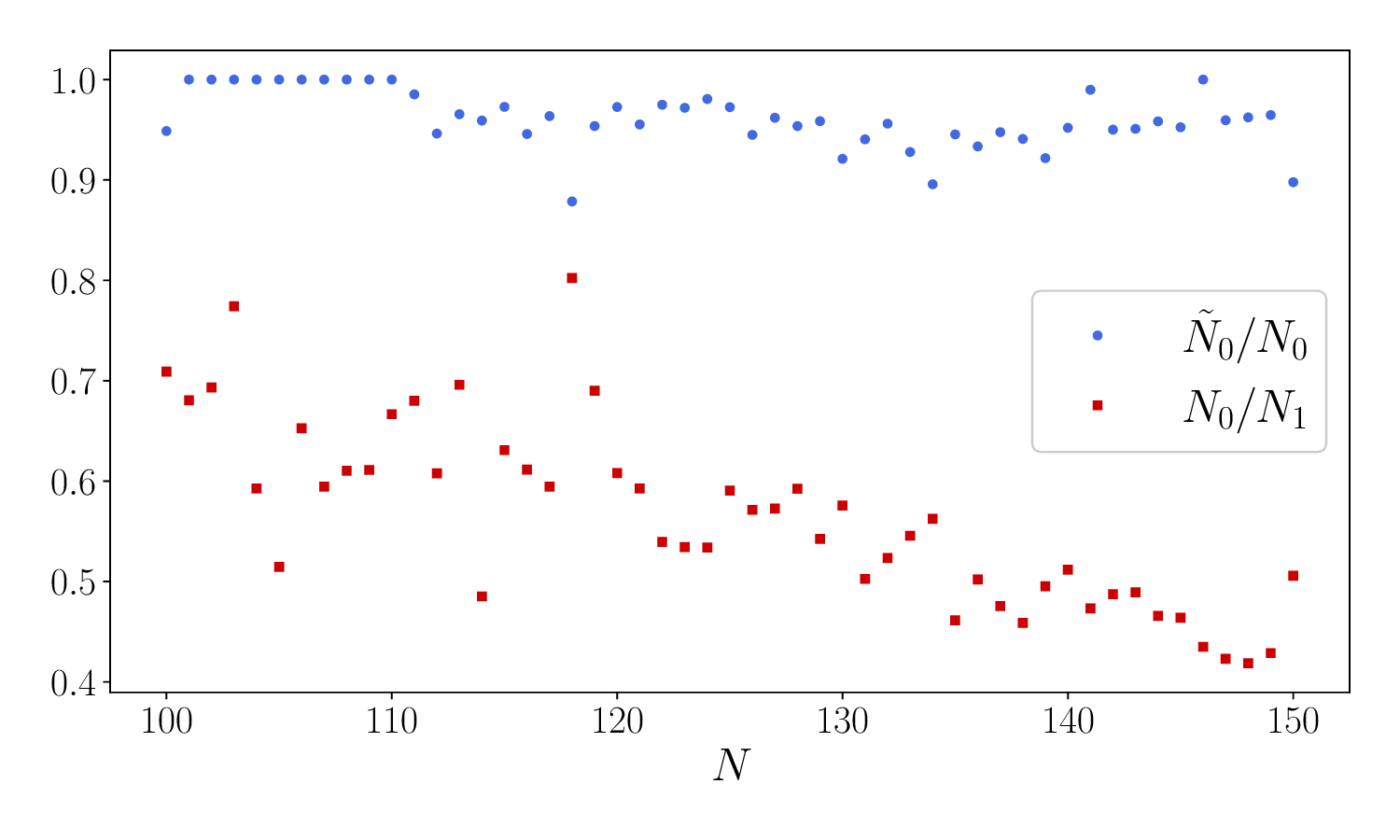}    
\caption{$\tilde{N}_0/N_0$ vs $N$ (blue curve) and $N_0/N_1$ (red curve).}
\label{Fig_ratio_s1_s0}
\end{center}
\end{figure}

To get started in our exploration we took advantage of our previous exploration for the Thomson problem of ref.~\cite{Amore25} and we used the independent
configurations for $s=1$ as ansatzes for the problem with $s=0$: once the potential is changed from Coulomb to logarithmic, the configurations cease to be local  minima of the new potential, but are typically quite close to a local minimum, thus converging quite fast to the new solution. 
The results are illustrated in Fig.~\ref{Fig_ratio_s1_s0}: the blue points represent the ratio between $\tilde{N}_0$, the number of local minima for $s=0$ obtained from minimization of local minima at $s=1$ and $N_0$, the total number of local minima at $s=0$; the red points correspond to the ratio between $N_0$ and $N_1$ (total number of local minima at $s=1$). Notice that $N_0$ and $N_1$ are obtained performing a full exploration of the energy landscape. 

Remarkably, this simple adaptation of the local minima for $s=1$ accounts for more than $90 \%$ of the total configurations at $s=0$ for $100 \leq N \leq 150$ (blue points), whereas the ratio $N_0/N_1$ ranges from $0.4$ (close to $N=150$) up to $0.8$ (close to $N=120$). 

\begin{figure}
\begin{center}
\includegraphics[width=10cm]{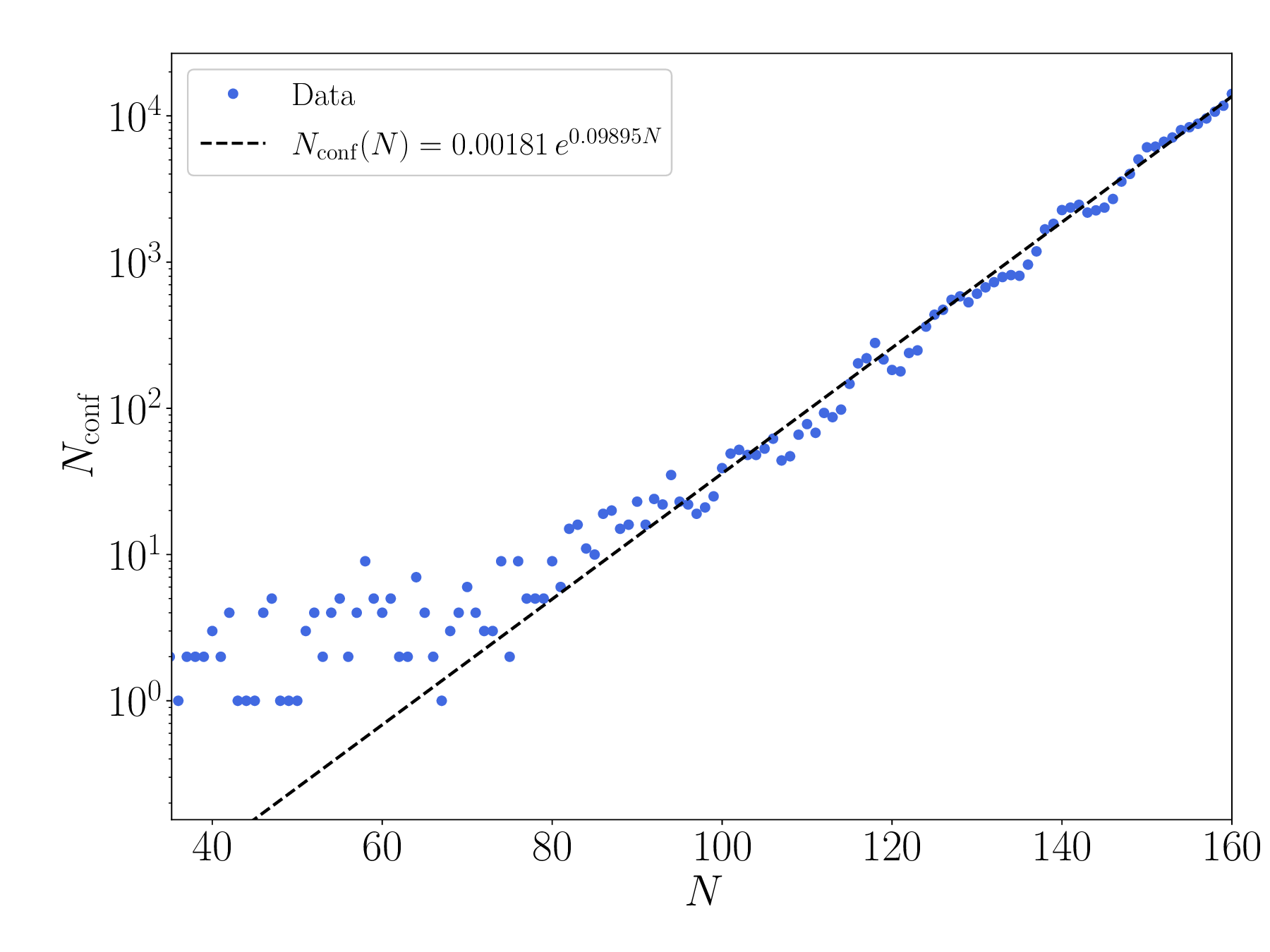}    
\caption{Number of local minima for the Thomson problem. The fit was done for values in the range $100 \leq N \leq 160$.}
\label{Fig_N_local}
\end{center}
\end{figure}

In Fig.~\ref{Fig_N_local} we plot $N_{\rm conf}$ for $10 \leq N \leq 160$ (blue dots in the plot) and the fit of the results for $100 \leq N \leq 160$
\begin{equation}
N_{\rm conf}(N)  = 0.00181 \times e^{0.09895\ N} \ .
\label{fit_nconf}
\end{equation}
 Two observations are in order: first, that the growth of $N_{\rm conf}$ that we observe is stronger that the one observed in  \cite{Calef15} and, second, that the growth   for the logarithmic potential is milder than the one found for the Coulomb potential~\footnote{Ref.~\cite{Calef15} does predict that the number of configurations for the logarithmic potential is smaller than for the Coulomb potential, but to less extent.}.    This last behavior  can be qualitatively understood if one realizes that larger values of $s$ correspond to shorter range potentials, where distant particles are less correlated between each other, thus reducing some constraints in forming an equilibrium configuration.

\begin{figure}
\begin{center}
\includegraphics[width=10cm]{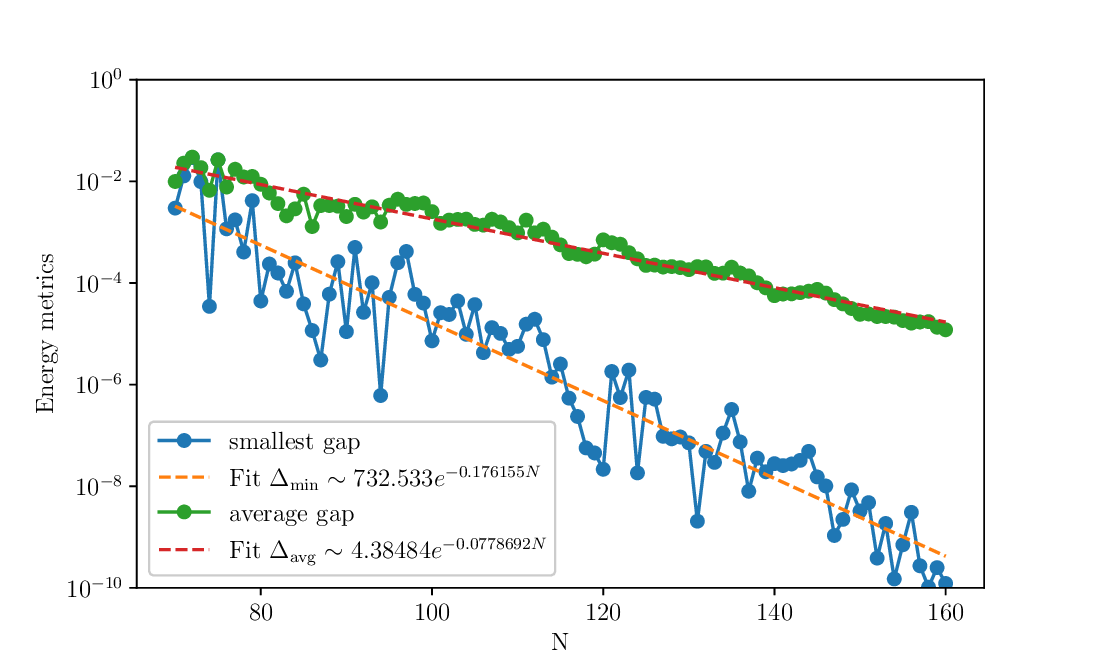}    
\caption{Smallest energy gap for configurations with $70\leq N \leq 160$.}
\label{Fig_smallest_gap_s0}
\end{center}
\end{figure}

In Fig.~\ref{Fig_smallest_gap_s0} we plot  the smallest energy   and average energy gaps for local minima configurations of the logarithmic potential, for 
$70 \leq N \leq 160$, in analogy to what we had done in \cite{Amore25} for the Coulomb potential.  

In the present case we find that the data are well  approximated by the fits
\begin{equation}
\begin{split}
\Delta_{\rm min} &= 732.533 \ e^{-0.176155 \ N} \\
\Delta_{\rm avg} &= 4.38484 \ e^{-0.0778692 \ N}
\end{split} \ ,
\end{equation}
confirming qualitatively our previous observations. In particular, for $N \gtrsim 400$ the average gap may become so small that would impair energy comparison working in double precision.

\begin{figure}
\begin{center}
\includegraphics[width=10cm]{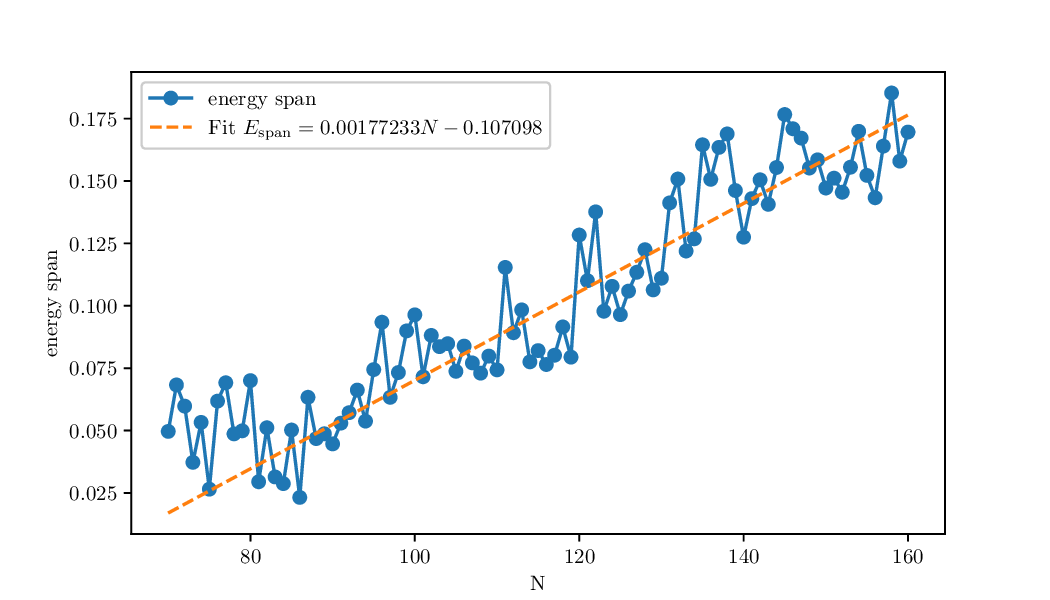}    
\caption{Energy span defined as $E_{\rm max}(N) -E_{\rm min}(N)$ for configurations with $70\leq N \leq 160$.}
\label{Fig_energy_span_s0}
\end{center}
\end{figure}

Similarly, the energy span in the present case is found to grow linearly with $N$ (see Fig.~\ref{Fig_energy_span_s0}) as
\begin{equation}
E_{\rm max}(N) - E_{\rm min}(N)  \approx 0.00177 N  - 0.107098 \ .
\end{equation}

\begin{figure}
\begin{center}
\includegraphics[width=10cm]{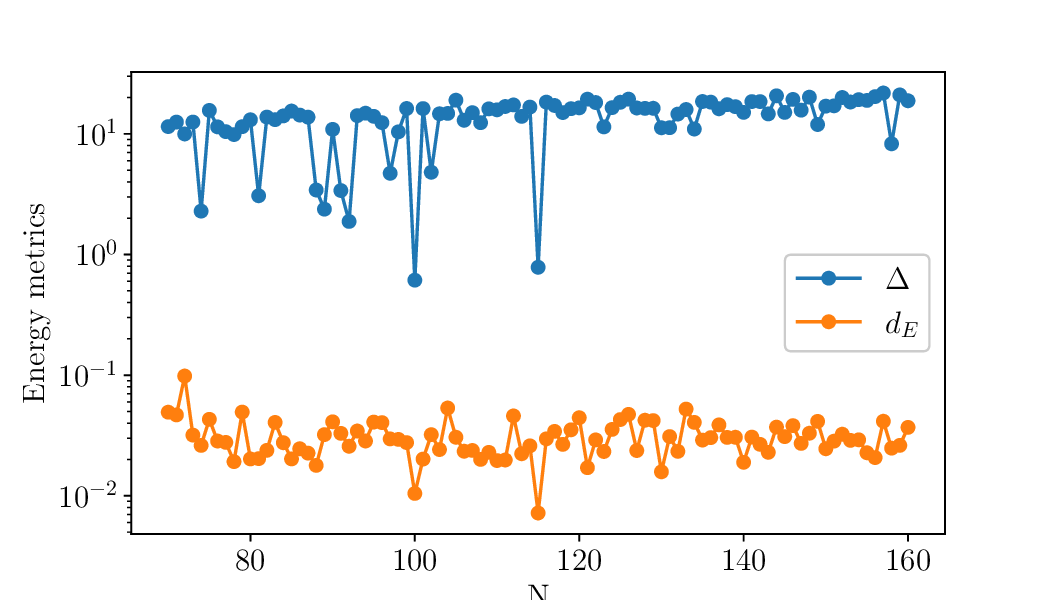}    
\caption{Smallest $\Delta$ and $d_E$ for  $20\leq N \leq 160$.}
\label{Fig_gap_comparison_s0}
\end{center}
\end{figure}

Another aspect of interest, is the possible occurrence of nearly degenerate configurations: in ref.~\cite{Amore25} we found  examples of these configurations  for the Coulomb potential  at $N=104$ and $N=114$; in Fig.~\ref{Fig_gap_comparison_s0} we carry out a similar analysis for the logarithmic potential, for $20 \leq N \leq 160$. In this case, however, we find that nearly degenerate solutions are less common than for the Coulomb potential.

\begin{figure}
\begin{center}
\includegraphics[width=12cm]{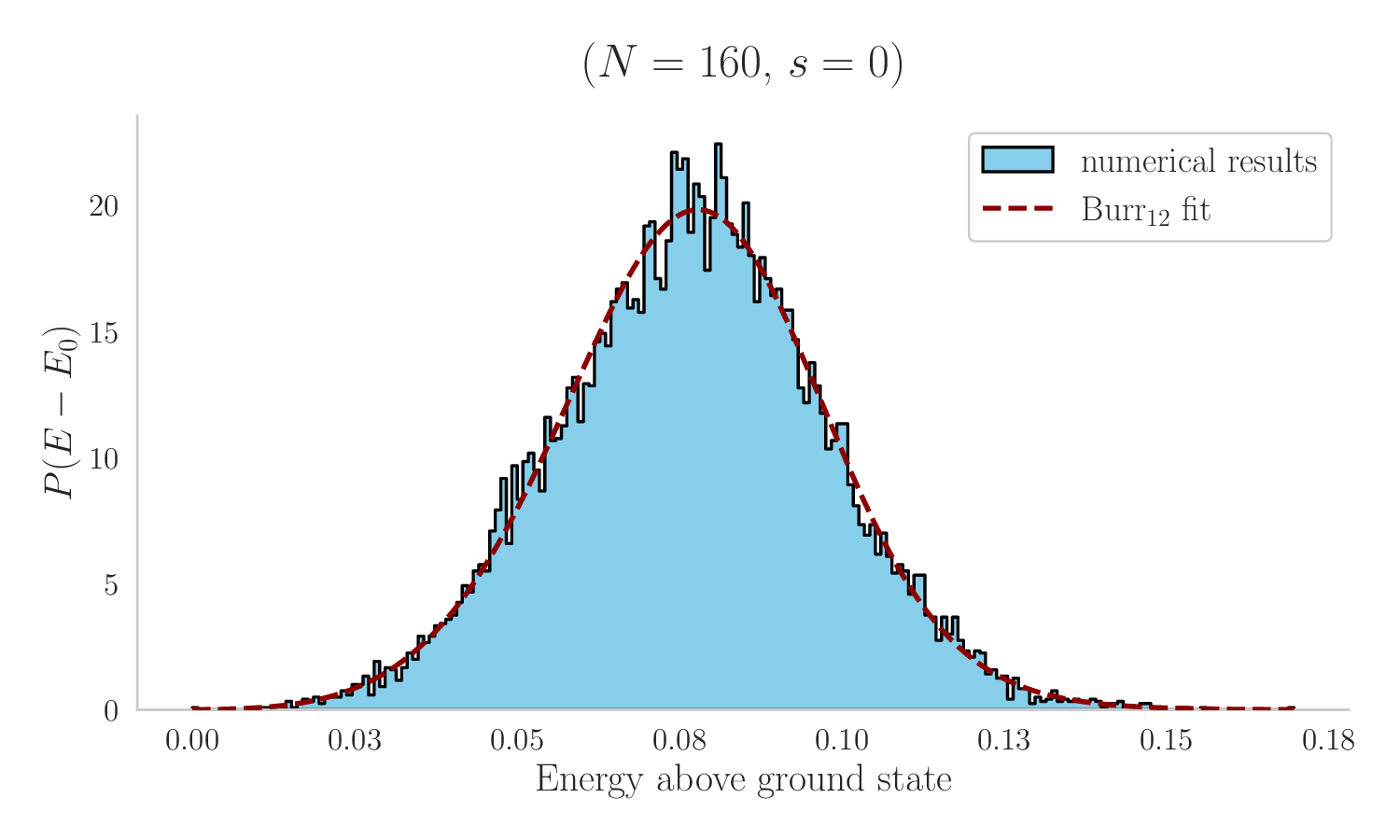}    
\caption{Histogram for the energies of independent configurations for $N=160$ corresponding to $f(x;c,k,\lambda) = \frac{ck}{\lambda}\left(\frac{x}{\lambda}\right)^{c-1}\left[1 + \left(\frac{x}{\lambda}\right)^{c}\right]^{-(k+1)}\quad\text{where}\quad c=5.49,\ k=4.24,\ \lambda=0.1213$ (using 200 bins).}
\label{Fig_energy_distribution_s0}
\end{center}
\end{figure}

\begin{figure}
\begin{center}
\includegraphics[width=12cm]{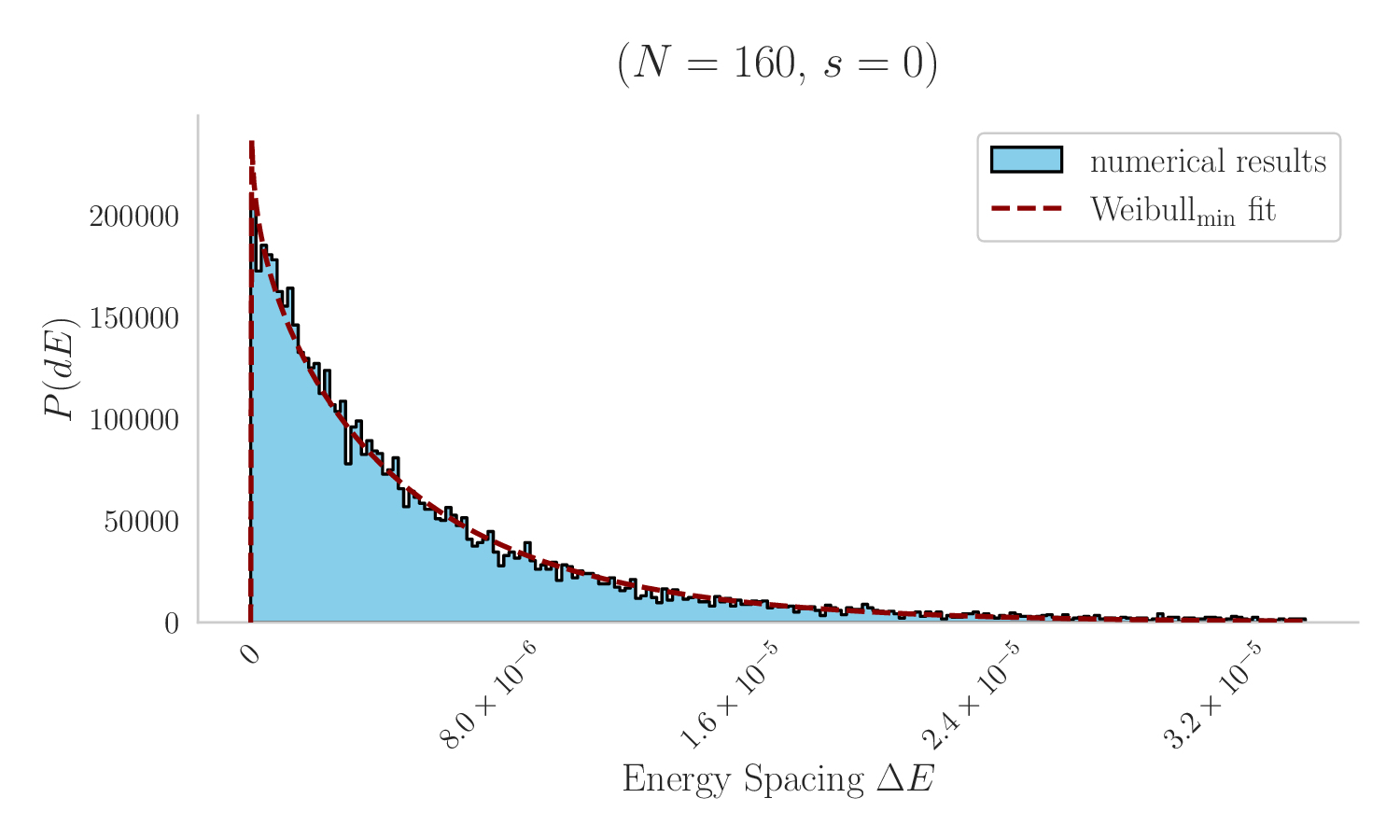}    
\caption{Histogram for the energy gaps (right)  for $N=160$ corresponding to $f(x;\lambda,k) = \frac{k}{\lambda}\left(\frac{x}{\lambda}\right)^{k-1}e^{-\left(\frac{x}{\lambda}\right)^k}\quad\text{where}\quad k = 0.932,\ \lambda = 5.499 \times 10^{-6}$ (using 200 bins)
}
\label{Fig_denergy_distribution_s0}
\end{center}
\end{figure}

Also for the logarithmic potential we find that the energies of the independent configurations appear to be distributed following a Burr {\rm XII} distribution~\cite{Sanchez19}.
For the case of $N=160$ points, displayed in  Fig.~\ref{Fig_energy_distribution_s0}, we find
\begin{equation}
f(x;c,k,\lambda) = \frac{ck}{\lambda}\left(\frac{x}{\lambda}\right)^{c-1}\left[1 + \left(\frac{x}{\lambda}\right)^{c}\right]^{-(k+1)}
\end{equation}
where $c=5.49$, $k=4.24$  and $\lambda=0.1213$ .

The energy gaps for $N=160$ are well described by  a Weibull distribution  (Fig.~\ref{Fig_denergy_distribution_s0}):
\begin{equation}
f(x;\lambda,k) = \frac{k}{\lambda}\left(\frac{x}{\lambda}\right)^{k-1}e^{-\left(\frac{x}{\lambda}\right)^k} \ , 
\end{equation}
where $k = 0.932$ and $\lambda = 5.499 \times 10^{-6}$.  As we notice in \cite{Amore25}, Weibull distribution is used to describe fragment
size distribution~\cite{Tenchov86,Fang93,Lu02,Brouwers16,Fanfoni24}.

\begin{figure}
\begin{center}
\includegraphics[width=14cm]{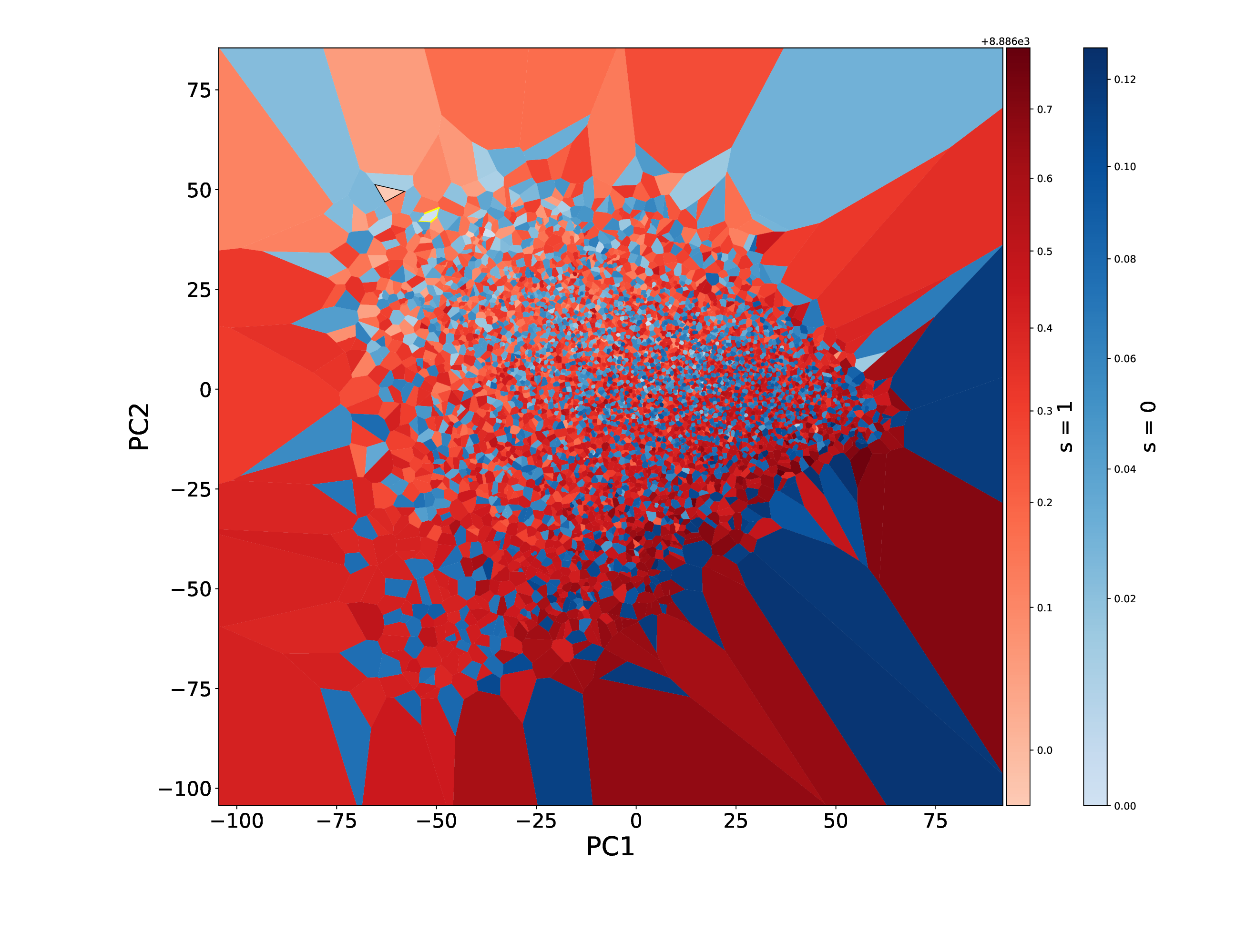}    
\caption{PCA for  local minima with 140 charges for the Coulomb (red colors) and logarithmic (blue colors)  potentials. The Voronoi cells with black and green borders mark the global minima
for the Coulomb and logarithmic potential respectively. }
\label{Fig_PCA_140_s0}
\end{center}
\end{figure}

\begin{figure}
	\begin{center}
		\includegraphics[width=12cm]{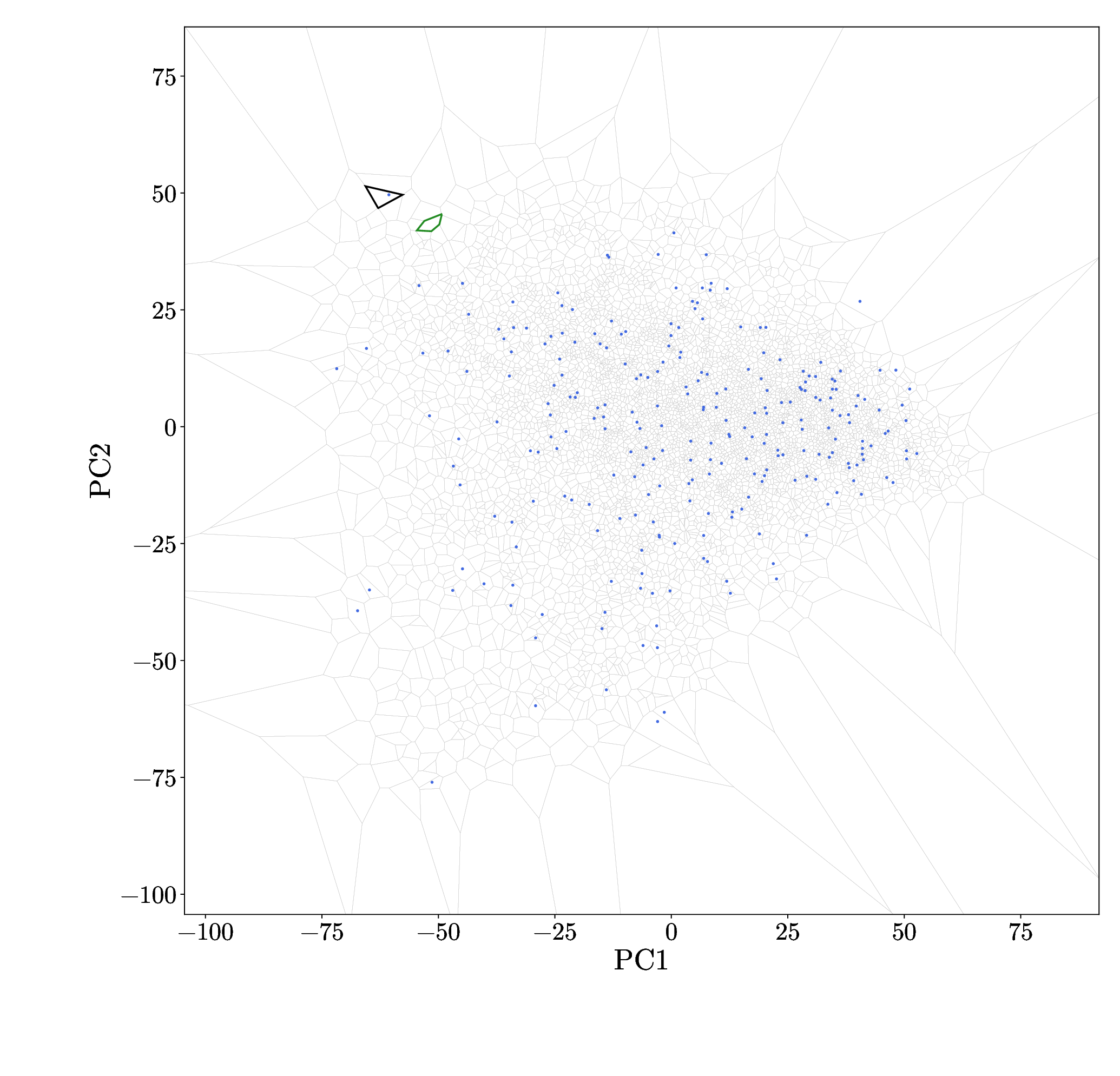}    
		\caption{PCA for  local minima with 140 charges: the blue dots correspond to local minima of the Coulomb potential that evolve to the global minima of the logarithmic potential upon minimization. The cells corresponding to the global minima of the Coulomb and logarithmic potentials have black and green borders, respectively.}
		\label{Fig_PCA_140_s0_collapse}
	\end{center}
\end{figure}

In analogy with what we had done in  ref.~\cite{Amore25}, in Fig.~\ref{Fig_PCA_140_s0}  we show  a PCA diagram for the configuration of $N=140$ points on the sphere.  
This diagram, however, contains the local minima of both potentials, logarithmic and Coulomb, with the Voronoi cells being visualized in a blue and red color scales respectively, with the specific color in each scale being related to the energy of the local minimum. The Voronoi cells corresponding to the global minima are visualized with a green (logarithmic) and black (Coulomb) border.

We observe that:
\begin{itemize}
\item the closeness of two different Voronoi cells is an indicator of the degree of similarity of the  two configurations;
\item there is a predominance of red (Coulomb) over blue (logarithmic) colors in this diagram, due to  the relative abundance of local minima of the Coulomb potential compared to those of the logarithmic potential;
\item the global minima for the two potentials are rather close to each other;
\end{itemize}

In  Fig.~\ref{Fig_PCA_140_s0_collapse} we plot the same diagram, in a different way: this time, the Voronoi cells are not colored, 
but small blue points are plotted in correspondence of  local minima of the Coulomb potential that evolve to the configuration of the global minimum of the logarithmic potential, upon minimization. 

It is  interesting to observe that 
\begin{itemize}
\item there is a large number of blue points (for $N=140$ there are $253$  cases -- out of $4434$  --  where the minima of $s=1$ have evolved to the global minimum of $s=0$) ;
\item the blue points are scattered over a rather large portion of the diagram and not just close to the global minimum for $s=0$ ;
\item the energies of the configurations corresponding to the blue points take values over a wide range of values;
\end{itemize}
At least for the modest values of $N$ that we are able to consider in the present paper, reaching the global minima  of the potential is relatively easy, confirming the observations of  \cite{Mehta16}.

\begin{figure}
\begin{center}
\includegraphics[width=8cm]{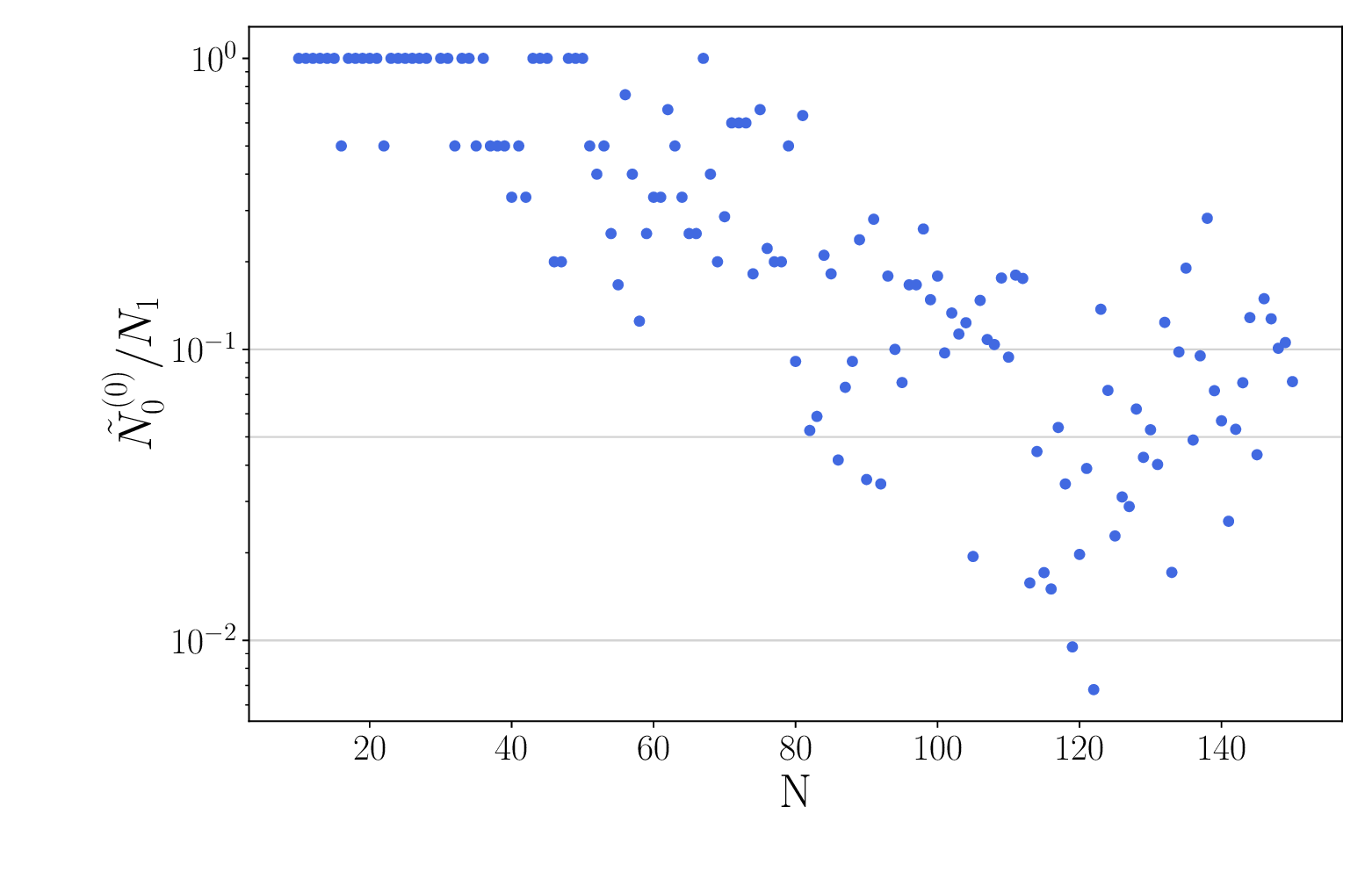}    
\caption{Fraction of local minimum configurations of $s=1$ that evolve to the global minimum of $s=0$ upon  minimization, versus $N$. The thin horizontal lines correspond to $0.01$, $0.05$ and $0.1$.}
\label{Fig_Ngs_vs_N}
\end{center}
\end{figure}

In Fig.~\ref{Fig_Ngs_vs_N} we plot the fraction of local minimum configurations of $s=1$ that evolve to the global minimum of $s=0$ upon  minimization, versus $N$:  with the only exception of $N=119$ and $122$, all values fall above the $0.01$ level. Keeping in mind the exponential growth of the number of configurations with $N$, this behavior reflects in a very large number of configurations that collapse to the global minimum for $s=0$. For example, for $N=149$ we have found 
$11766$ local minimum configurations for the Coulomb potential and $1242$ of them collapse to the global minimum of the logarithmic potential upon minimization.

\begin{figure}
\begin{center}
\includegraphics[width=12cm]{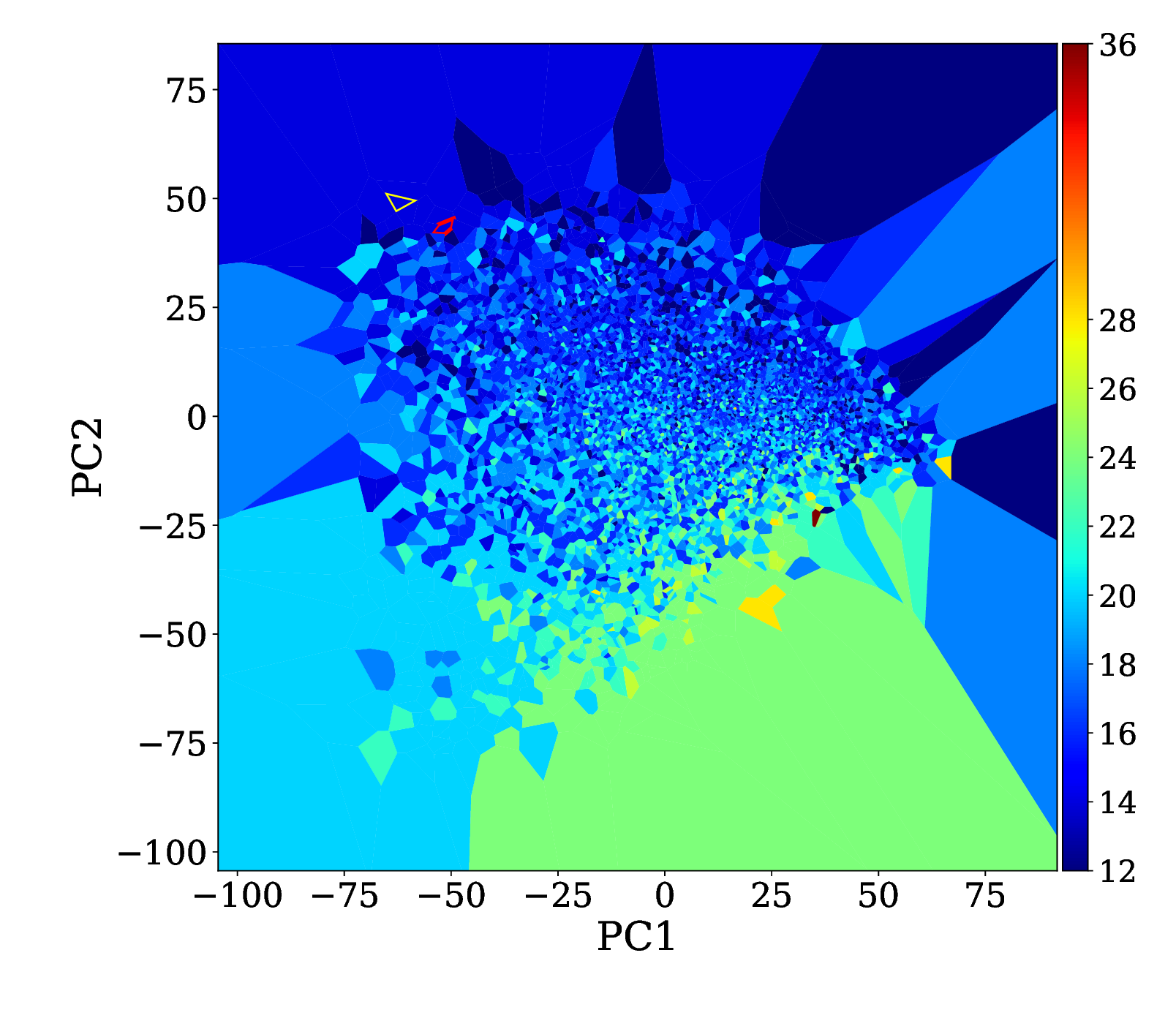}    
\caption{PCA for  local minima with 140 charges: the color of the Voronoi cells reflects the value of $\mathcal{Q}$. The borders of the cells corresponding to the global minima for $s=1$ and $s=0$ are highlighted in yellow and red, respectively.}
\label{Fig_PCA_140_defects}
\end{center}
\end{figure}

Finally, we have produced a different version of the PCA diagram, Fig.~\ref{Fig_PCA_140_defects}, this time coloring the Voronoi cells according to the sum of the absolute values of the topological charges: 
Euler's theorem requires that the total topological charge adds up to $Q = \sum_{i=1}^N q_i =  12$, but it does not constrain
\begin{equation}
\mathcal{Q} = \sum_{i=1}^N |q_i|  \ ,
\end{equation}
where $q_i$ is the topological charge of the $i^{th}$ Voronoi cell. The lowest value that $\mathcal{Q}$ can take is $12$, corresponding to having $12$ pentagonal cells (with the remaining cells being hexagonal); larger values of $\mathcal{Q}$ correspond to having additional defects added, without altering the total topological charge. At $N=140$ considered in this figure,  the global minima  correspond to  $\mathcal{Q}  = 14$, signaling that the Voronoi diagrams of these configurations  contain $13$ pentagons and $1$ heptagon. 

At larger $N$ it is well known that the Voronoi diagrams of the global minima tend to display complex defect structures (such as for the beautiful "rosettes" found in \cite{Wales06,Wales09} for the Thomson problem): an exploration of the energy landscape in that case would  produce a much richer diagram, with the global minima now corresponding  to much larger values of $\mathcal{Q}$~\footnote{Of course, a full exploration of the energy landscape is impossible in regimes with $N \gtrsim 200$, due to the exponential growth of the number of configurations.}.

\begin{figure}
\begin{center}
\includegraphics[width=6cm]{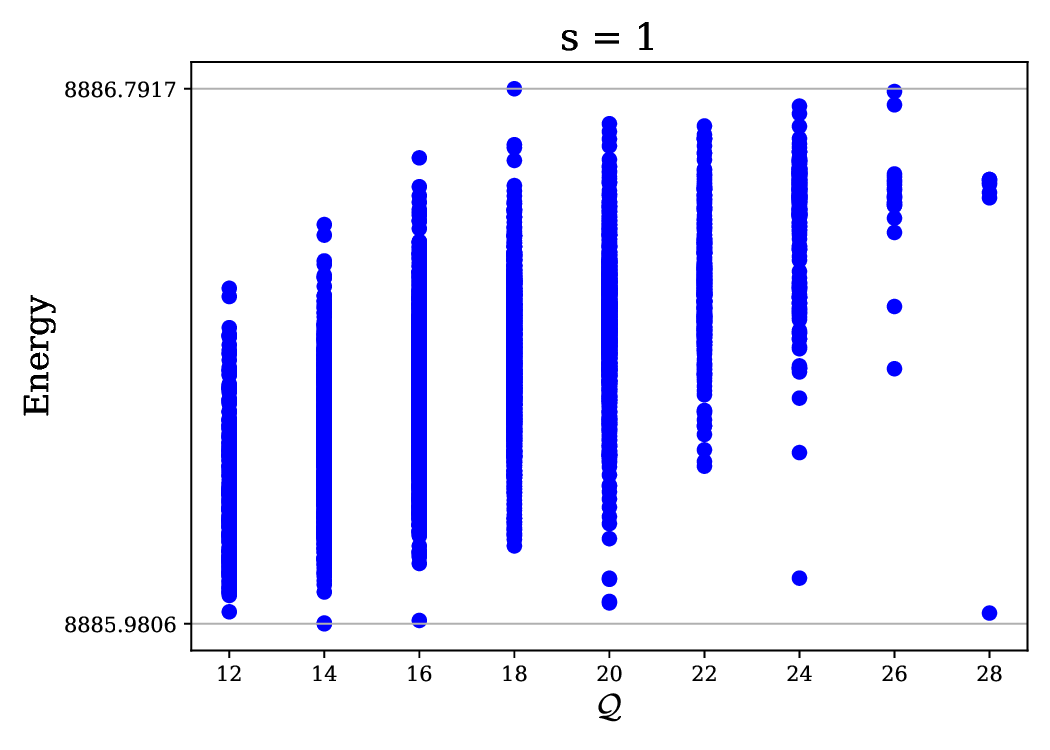}    
\includegraphics[width=6cm]{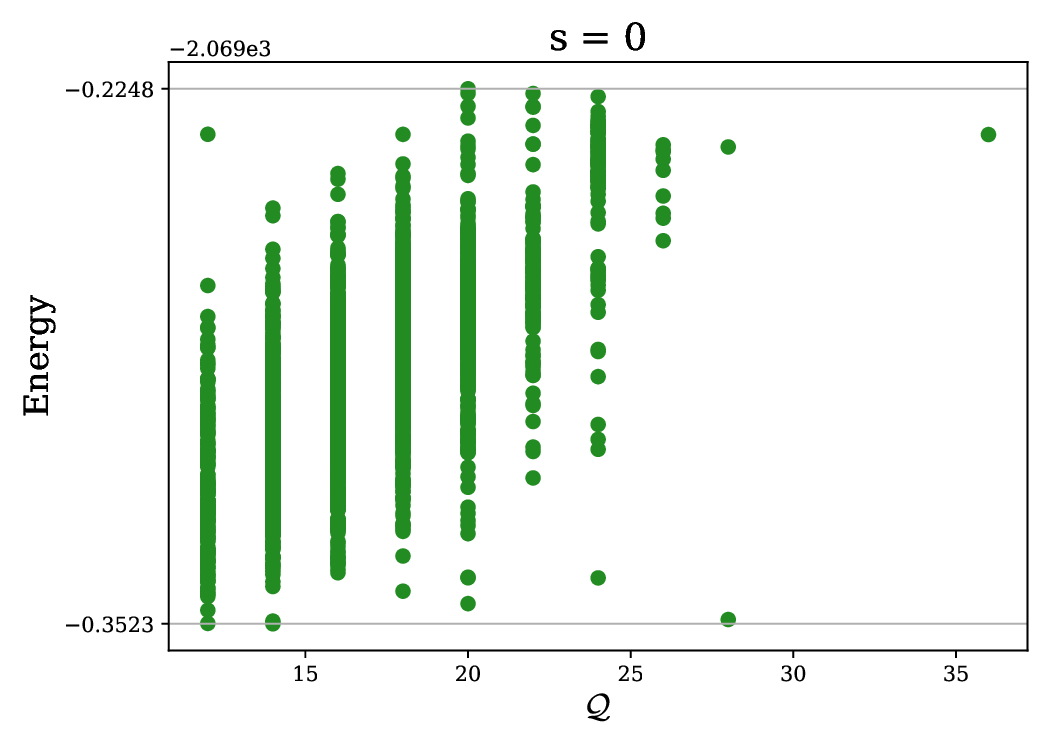}    
\caption{Energy of the configurations of $N=140$ points vs $\mathcal{Q}$ for the Coulomb (left plot) and the logarithmic (right plot) potentials.}
\label{Fig_Q_vs_E}
\end{center}
\end{figure}

\begin{figure}
\begin{center}
\includegraphics[width=10cm]{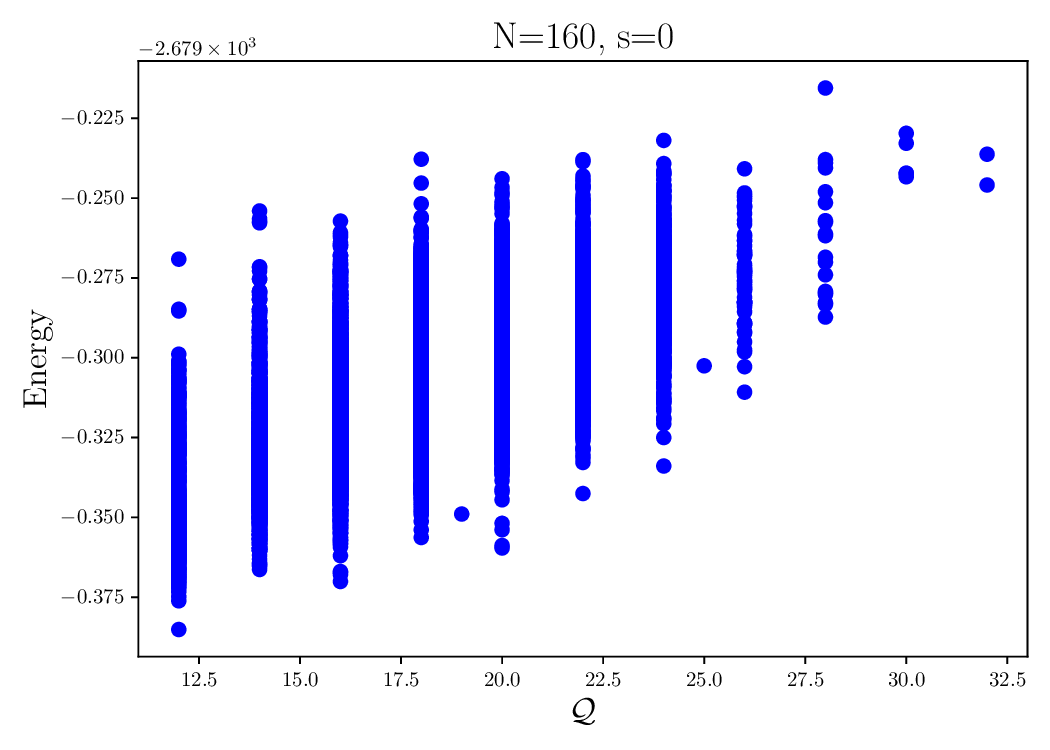}    
\caption{Energy of the configurations of $N=160$ points vs $\mathcal{Q}$ for the logarithmic potential.}
\label{Fig_Q_vs_E_160}
\end{center}
\end{figure}

\begin{figure}
\begin{center}
\includegraphics[width=3cm]{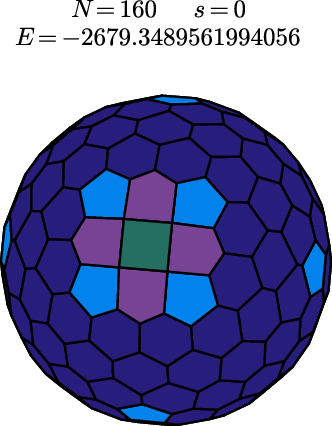}    \hspace{3cm}
\includegraphics[width=3cm]{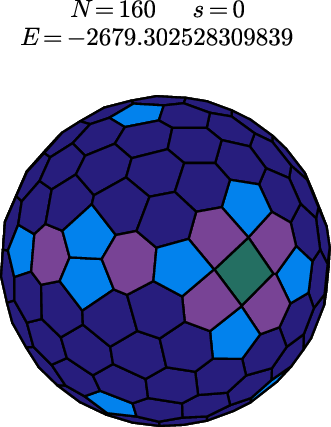}    
\caption{Configurations of $N=160$ points with $\mathcal{Q} = 19,25$}
\label{Fig_160_conf}
\end{center}
\end{figure}

In Figs.~\ref{Fig_Q_vs_E} we plot the energies of the configurations of $N=140$ points for $s=1$ (left plot) and $s=0$ (right plot) as functions of $\mathcal{Q}$.  Notice that  all the configurations found correspond to even  values of $\mathcal{Q}$: to obtain odd values of $\mathcal{Q}$ one would need to have a topological defects containing  a square or octagonal cell. Such situation occurs for larger $N$: in Fig.~\ref{Fig_Q_vs_E_160}  we consider the case of $N=160$ points interacting via the logarithmic potential: in this case we find  two configurations corresponding to $\mathcal{Q}=19,25$. These "exotic" configurations are shown in Fig.~\ref{Fig_160_conf}.

\subsection{Stationary states}
\label{sec:stat_states} 

Following the approach that we have introduced in \cite{Amore25} we can look for the stationary states of the logarithmic potential by considering the auxiliary problem with the effective potential $\mathcal{V}$ defined in eq.~(10) of ref.~\cite{Amore25}.

In this framework, the exploration of the solution landscape corresponds to finding all the global minima of $\mathcal{V}$, which are stationary configurations of the original potential $V$.   This exploration, however, is not straightforward, even for rather modest values of $N$,   due to the large number of stationary configurations.  In fact, we believe that our results constitute the first attempt to study the solution landscape for the logarithmic potential.

\begin{figure}
\begin{center}
\includegraphics[width=10cm]{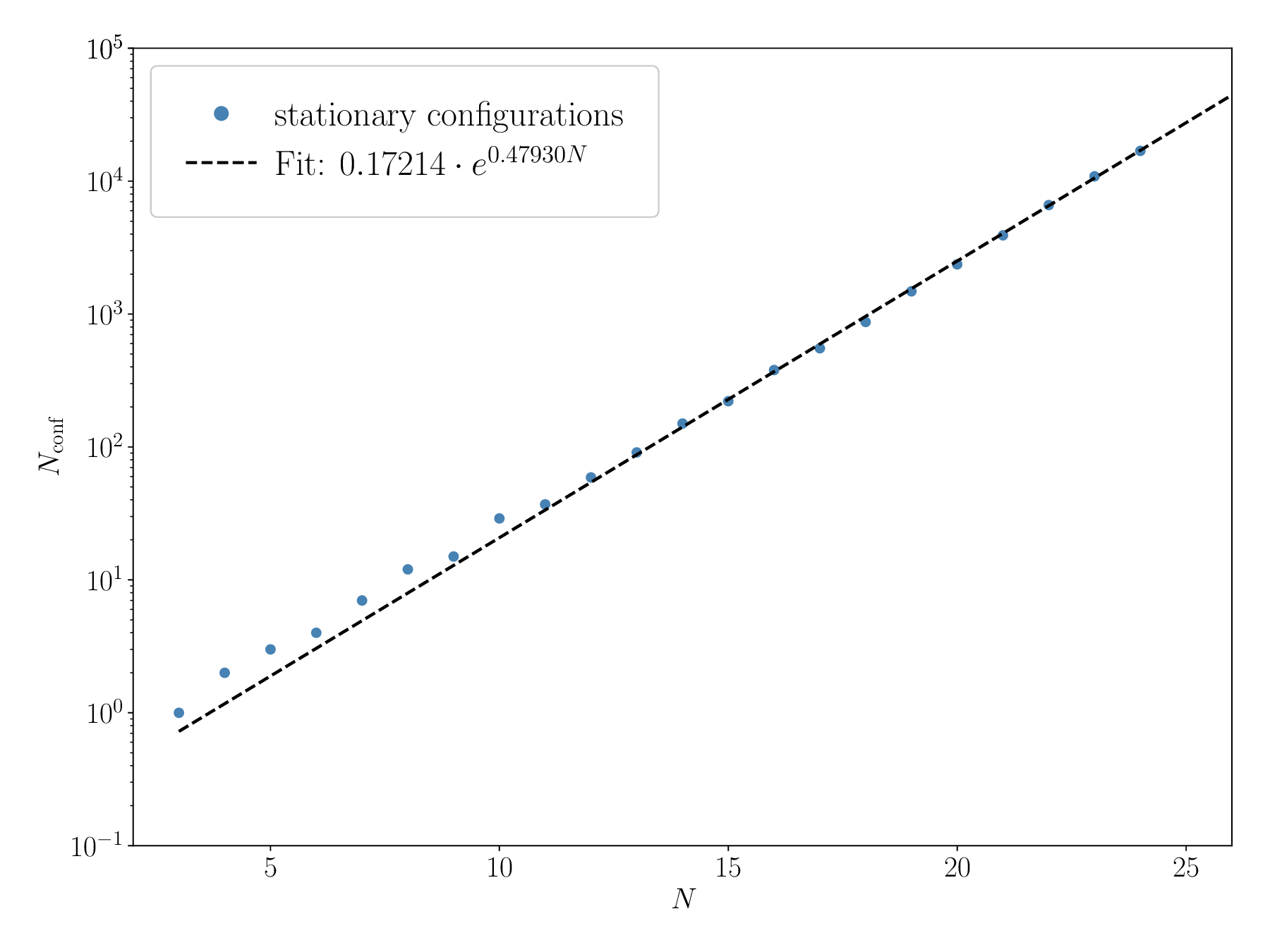}    
\caption{Number of stationary configurations for the logarithmic potential.}
\label{Fig_N_stationary_s0}
\end{center}
\end{figure}

As for the Coulomb problem that we previously studied, we have conducted  a large number of numerical experiments for $2 \leq N \leq 24$:
these results are summarized in Fig.~\ref{Fig_N_stationary_s0} where we report the number of stationary configurations found vs $N$ for the different cases studied (blue dots); the dashed line in the plot  is the exponential fit over the range $15 \leq N \leq 24$
\begin{equation}
N_{\rm conf} \approx 0.17214 \times e^{0.4793  \ N} \ .
\end{equation}

As for the case of local minima, we find that the exponential growth of the number of stationary configurations is milder than the one we observed for the case of the Coulomb potential in \cite{Amore25}. This result is not surprising and it confirms the expectation that shorter range potentials have a richer solution space.

\begin{figure}
\begin{center}
\includegraphics[width=10cm]{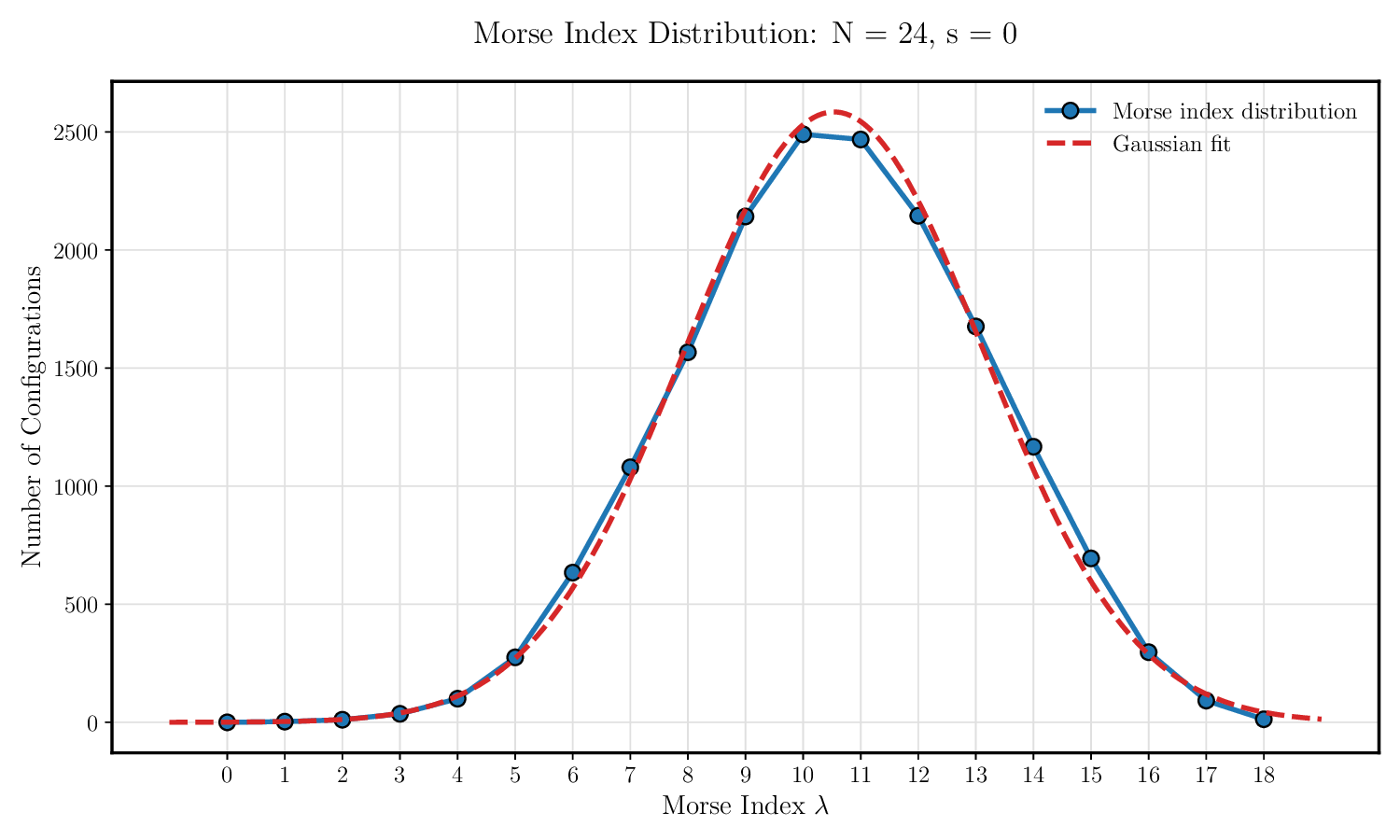}
\caption{Number of stationary configurations for the logarithmic potential.}
\label{Fig_N_stationary_vs_index_s0}
\end{center}
\end{figure}

In Fig.~\ref{Fig_N_stationary_vs_index_s0} we plot the number of stationary configurations of the logarithmic potential as function of the Morse index for $N=24$: the observed behavior is very similar to the one previously found for the Coulomb potential (with the peak of the distribution being slightly smaller in the present case).

\begin{figure}
\begin{center}
\includegraphics[width=10cm]{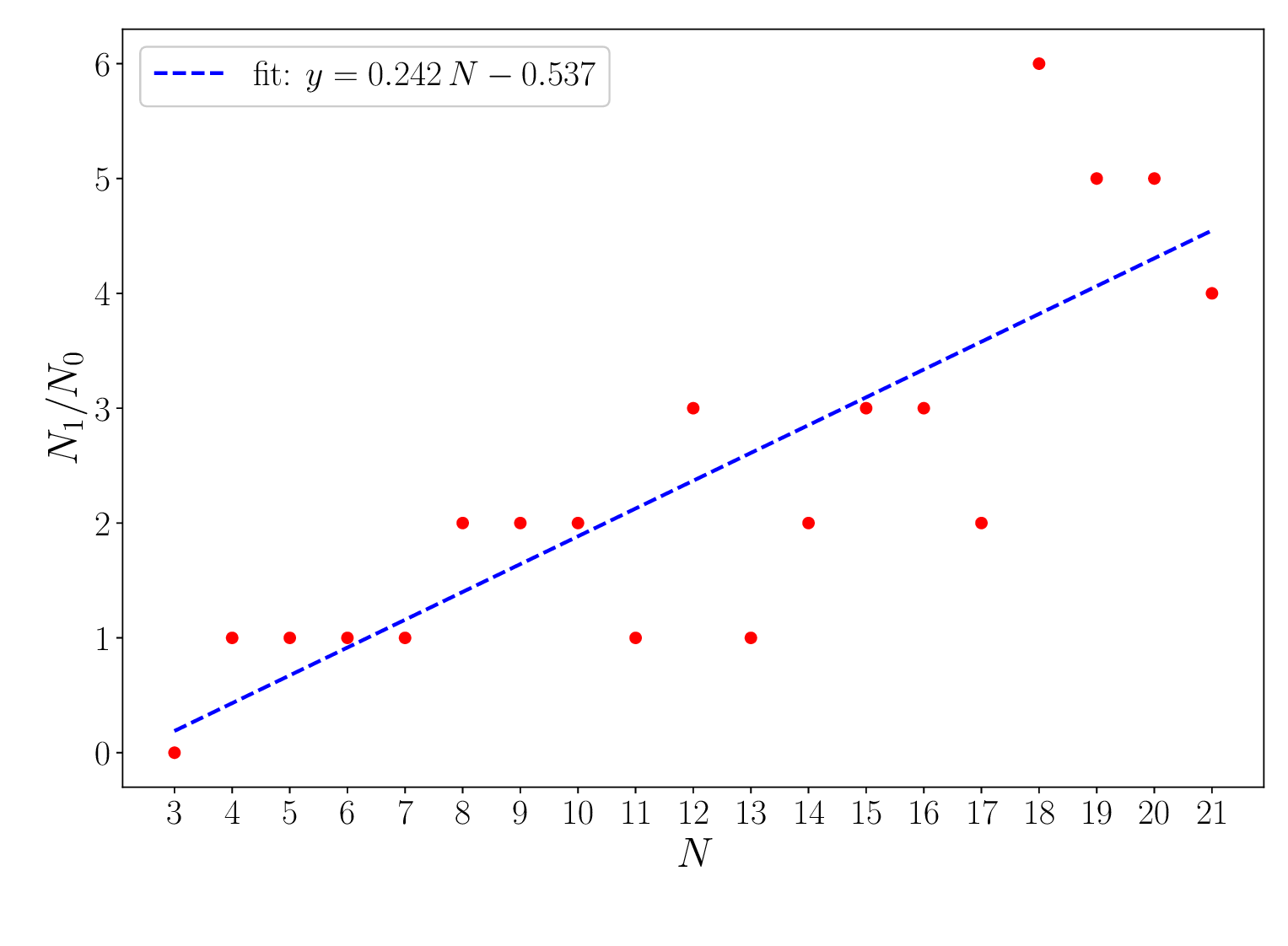}    
\caption{ Ratio between the number of transition points and local minima, $N_1/N_0$,  as a function of $N$.}
\label{Fig_ratio_n1_n0}
\end{center}
\end{figure}

Finally, in Fig.~\ref{Fig_ratio_n1_n0} we plot the ratio between the number of transition points and local minima, $N_1/N_0$,  as a function of $N$, for $N \leq 21$. The behavior is quite similar to the one we previously observed for the Coulomb potential in \cite{Amore25} and in accordance with the prediction of ref.~\cite{Wales03b}.

\section{Conclusions}
\label{sec:concl} 

We have performed an in--depth exploration of the energy and of the solution landscapes for point charges on a sphere interacting 
via a logarithmic potential.

Many features that were already observed for the Coulomb potential are also found here; in particular
\begin{itemize}
\item the number of local minima and of stationary configurations grows exponentially with $N$; 
\item the energy gap (both minimum and average) decay exponentially with $N$, whereas the energy span grows linearly;
\item the distributions of the energy and of the energy gaps are well described by a Burr$_{12}$ and a Weibull$_{min}$ distributions;
\item the stationary configurations at a given $N$ appear to follow a Gaussian distribution in the Morse index, 
\end{itemize}

We also find that the number of independent minima and, more in general, stationary points, is smaller for the logarithmic potential than for the Coulomb potential: this behavior can be understood as a consequence of the longer range nature of the logarithmic potential. The PCA for the local minima configurations for $s=0$ and $s=1$ of $N=140$ charges on the sphere shows that the minima of the logarithmic potential
tend to be distributed over the landscape and immersed, so to say, in a background of minima of the Coulomb potential (see Fig.~\ref{Fig_PCA_140_s0}). Moreover, the global minima of the two problems are quite close (and hence similar).  One could ask
whether this picture is modified at larger $N$, because of its relevance, among other things, to Smale's seventh problem~\cite{Smale98}: 
in the present case we find that the global minimum of the logarithmic potential is well connected with regions of the landscape which are neither close or within a small energy window. 

For the cases studied in our paper, with the exceptions of $N=119$ and $122$, at least  $1\%$ of the local minima of the Coulomb potential collapse to the global minimum of the logarithmic potential, upon minimization. If this feature survives at larger $N$, it could inspire a strategy to look for the global minimum of the problem, in terms of a summary exploration of the landscape of a shorter range problem.   An in--depth exploration of the energy landscape, as the one done here, however, becomes literally impossible even for values of $N$ which are slightly larger than the one we have considered here, due to the massive exponential growth of the number of configurations. Alternative approaches, applicable to much larger $N$, must be considered in  this case. We plan to explore these aspects in future works.

\section*{Data availability}

The data produced in this work are available for download at \href{https://zenodo.org/records/17923310}{Zenodo}.

\section*{Acknowledgements}
 The  research of P.A.  is supported by Sistema Nacional de Investigadores (M\'exico). 
 The work of R. R. is supported by a KIAS Individual Grant with number QP094602 at
 Korea Institute for Advanced Study.
This work uses computational resources supported by the Center for Advanced Computation at Korea Institute for Advanced Study

\end{document}